\documentclass{amsproc}
\usepackage{amssymb,amstext}
\newtheorem{theorem}{Theorem}[section]

\theoremstyle{definition}

\newtheorem{setup}[theorem]{Set up}

\theoremstyle{remark}

\numberwithin{equation}{section}
  
  \newcommand{\Real}{\mathbb R}

  \newcommand{\Z}{\mathbb Z}

  \newcommand{\erz}[1]{\langle{#1}\rangle}

 \newcommand{\rk}{\mbox{\rm rank}}
 
 \newcommand{\T}{\mathcal T}

\newcommand{\cprod}{\rtimes}
\newcommand{\cprof}{\rtimes}

\newcommand{\W}{{\mathcal W}}
\newcommand{\JJ}{\mathcal P}

\newcommand{\dpe}{{\dim \! V}}

\begin{document}
\title{Cohomology groups for projection point patterns}

\author{Alan Forrest}

\email{stmt8027@bureau.ucc.ie}
\thanks{Invited talk presented by Johannes Kellendonk on the ICMP~2000}
\thanks{The first author was supported by the Norwegian Research Council,
the British Council and the EC Network "Non-commutative Geometry"}

\author{John Hunton}
\address{
The Department of Mathematics and Computer Science, University of
Leicester, Leicester, LE1 7RH, England}

\email{jrh@mcs.le.ac.uk}
\thanks{The second author was supported by the Norwegian Research Council
and the British Council}

\author{Johannes Kellendonk}
\address{School of Mathematics, Cardiff University, Cardiff CF2 4YH, Wales}

\email{kellendonkj@cf.ac.uk}
\thanks{The last author was supported by the Sonderforschungsbereich 288}

\date{September 30, 2000}

\begin{abstract}
Aperiodic point sets (or tilings) which can be obtained by the
method of cut and projection from higher dimensional periodic sets
play an important role for the description of quasicrystals. Their
topological invariants can be computed using the higher
dimensional periodic structure.  We report on the results obtained
for the cohomology groups of projection point patterns
supplemented by explicit calculations made by F.~G\"ahler for many
well-known icosahedral tilings.
\end{abstract}

\maketitle

\bibliographystyle{amsalpha}

\section{Introduction}

Among the basic input data for the description of an aperiodic
solid is the point set $\T$ of its equilibrium atomic positions.
It is a discrete subset of $\Real^d$ which enjoys a number of
properties depending on the particular nature of the solid
\cite{BHZ}. For systems which show diffractive behaviour, like
quasicrystals, one expects $\T$ to have certain repetivity
properties: e.g.\ $r$-patches, which are intersections of
$r$-balls with $\T$, repeat in a relatively dense way. $\T$ is
then a more or less regular pattern. For ideal quasicrystals there
is a preferred construction for such patterns: by projection out
of a higher dimensional periodic structure. These are the
projection point patterns we are interested in here.

We consider topological invariants for point patterns of
$\Real^d$. These invariants are defined as the cohomology groups
of the pattern groupoid. For projection point patterns matters
simplify enormously because the pattern groupoid is equivalent to
a groupoid defined by a Cantor dynamical system $(X,\Z^d)$ which
can be explicitly described. As a consequence, we only have to
deal with the cohomology of the group $\Z^d$ with coefficients in
the integer valued continuous functions $C(X,\Z)$. Furthermore,
these cohomology groups are isomorphic to the (unordered)
$K$-groups of the $C^*$-algebra associated with the pattern
groupoid.

The invariants yield an important step towards the classification
of point patterns. Moreover, the invariants play a role for the
labelling of the gaps in the spectrum of a Hamiltonian describing
the particle motion in the solid. The $C^*$-algebra of the pattern
groupoid is the algebra of observables in the tight binding
representation. The tight binding Hamiltonian belongs to it and so
do its spectral projections associated with gaps, i.e.\
projections on (generalized) eigenstates of all energies up to the
gap.  The $K_0$-class of such a projection furnishes a label for
the gap which is stable under perturbations. An additional input
from physics is a trace on the $C^*$-algebra coinciding in the
physically relevant representation with the trace per unit volume.
It induces a homomorphism from the $K_0$-group of the algebra to
$\Real$ and the values of the integrated density of states on gaps
belong to its image. This theory is due to Bellissard \cite{Be2}.
The present results do not yet complete the gap-labelling because
they do not contain information about the order and the image of
tracial states. For recent results in the latter direction see
\cite{BKL00}.

In the coming section we briefly describe the hull-construction of
a point set. After that we link this construction with the
non-commutative approach using $C^*$-algebras and $K$-theory. In
Section~4 we introduce the point sets on which we focus our
attention here, presenting the general results in Section~5 and
the examples in Section~6. The results of Section~5 (together with
the first example of Table~1) are taken from \cite{FHKmemoir} to
which we refer for all proofs and missing explanations\footnote{A
projection method pattern as in \cite{FHKmemoir} is a little bit
more general than a projection point pattern in that it allows for
additional decorations.}. The remaining examples have been
computed by Franz G\"ahler.

\section{Hull of a point set and its dynamical system}
Point sets in $\Real^d$ give rise to topologically highly
non-trivial spaces and dynamical systems by means of the so-called
hull construction: Let $B_r$ be the ball of radius $r$ around the
origin $0\in\Real^d$, $\partial B_r$ its boundary, and
$B_r(\T):=(B_r\cap \T)\cup\partial B_r$ called the $r$-patch of
the set $\T\subset\Real^d$. Then
$$D(\T,\T')=\inf\{\frac{1}{r+1}|d_H(B_r(\T),B_r(\T'))<\frac{1}{r}\},$$
where $d_H$ is the Hausdorff metric, is a metric on the set of all
closed subsets of $\Real^d$.
Thus two sets are close if they coincide on a large ball around
the origin up to a small discrepancy. We are interested in the
topology defined by that metric. The group $\Real^d$, acting on a
subset of $\Real^d$ by translation, acts continuously. The
(continuous) hull of $\T$ is the $D$-completion $M\T$ of the orbit
of $\T$ under the action of $\Real^d$. This action extends to the
closure and so we define the dynamical system $(M\T,\Real^d)$
associated to $\T$. This dynamical system is the starting point
for many investigations, see e.g.\
\cite{Rudolph89,RadinWolff92,AP}.

The space $M\T$ is compact under very general conditions, in
particular if $\T$ is a Delone set. But it may be quite
complicated. It consists of orbits of point sets each one being
homeomorphic to a copy of $\Real^d$. The closed subset
$\Omega\T=\{T\in M\T|0\in T\}$ intersects each orbit and this
intersection is transversal in the sense that very small
translations move any point of $\Omega\T$ outside it. If the
dimension $d$ is $1$ we can reduce by Poincar\'e's construction
the continous dynamical system $(M\T,\Real)$ to a discrete one
$(\Omega\T,\Z)$ without loss of topological information (the first
return map yielding the action of $\Z$). If $d$ is larger we
cannot in general find an action of $\Z^d$ on $\Omega\T$ which
generalizes Poincar\'e's construction but have to work with an
$r$-discrete groupoid (the pattern groupoid) instead. Projection
point patterns, however, fall into the category of point sets for
which exist another transversal $X$ such that $(M\T,\Real^d)$ can
be reduced to a discrete dynamical system $(X,\Z^d)$ without loss
of topological information. Another way of putting this is that
$M\T$ is the mapping torus of $(X,\Z^d)$, i.e.\ $M\T =
X\times\Real^d/\sim$ where $(x,y+a)\sim(\alpha_a(x),y)$ for all
$a\in\Z^d$ and $\alpha$ denotes the $\Z^d$ action.

\section{$C^*$-algebras, $K$-groups and cohomology}
Topological dynamical systems give rise to $C^*$-algebras by the
crossed product construction. In our context we have two such
algebras, $C(M\T)\cprof \Real^d$ and $C(X)\cprod\Z^d$. They are
strongly Morita equivalent
and therefore they have isomorphic $K$-theory (up to scale). As an
aside we mention that there is yet another construction of a
$C^*$-algebra from a point set $\T$, the groupoid $C^*$-algebra of
the pattern groupoid. This $C^*$-algebra, which exists for rather
general point sets, is strongly Morita equivalent to $C(M\T)\cprof
\Real^d$ as well. It has the interpretation as algebra of
observables for the tight binding approximation, for a review
see \cite{KellendonkPutnam00}.

The mere existence of the transversal $X$ allows one to connect
the $K$-theory with the cohomology of the group $\Z^d$. By Connes'
Thom isomorphism \cite{Bla}, the $K$-groups of the crossed product
$C(X)\cprod\Z^d$ are isomorphic to those of the continuous hull,
\begin{equation}\label{eq1}
K_*(C(X)\cprod\Z^d)\cong K_{*-d}(C(M\T)).
\end{equation}
A cofiltration of
$M\T$ gives rise to a spectral sequence whose $E_2$-term is
isomorphic to $H^*(\Z^d,K_*(C(X)))$, the cohomology of $\Z^d$ with
coefficients in $K_*(C(X))$. Forrest and Hunton \cite{FoHu} have
established that, if $X$ is homeomorphic to the Cantor set, the
spectral sequence collapses at the $E_2$-term and
\begin{equation}\label{KH}
K_{i}(C(M\T))\cong \bigoplus_{j}H^{2j+i}(\Z^d,C(X,\Z)).
\end{equation}
For projection point patterns, $X$ is indeed homeomorphic to the
Cantor set and the machinery of spectral sequences proves also to
be useful in explicitly computing the cohomology-groups. Below we
present the results of a refined analysis for canonical projection
method patterns or tilings \cite{FHKmemoir}. We note, however,
that this approach fails to give information about the order on
$K_0$ and the ranges of tracial states.

\section{Projection point patterns}

Projection point patterns are obtained by cut and projection from
higher-dimensional periodic structures, see e.g.\ \cite{DuKa}. Let
$\Lambda$ be a rank $N$ lattice spanning $\Real^N$ and $E$ be a
linear subspace intersecting $\Lambda$ only trivially. Let
$E^\perp$ be a complimentary subspace and denote by
$\pi:\Real^N\to E$ and $\pi^\perp:\Real^N\to E^\perp$ the
projections with kernel $E^\perp$ and $E$, respectively. Finally
let $K\subset E^\perp$ be a compact subset which is the closure of
its interior, called the acceptance domain.
$$P(K):=\{\pi(x)|x\in\Lambda, \pi^\perp(x)\in K\}$$ is the
projection point pattern defined by the data $(\Lambda,E,K)$.
There is a canonical choice for $K$, namely the projection of the
unit cell for $\Lambda$ under $\pi^\perp$, but also fractal $K$ is
of interest (although we will have nothing to say about this
case). The euclidian closure of $\pi^\perp(\Lambda)$ can be
written as $V+\Delta$, where $V$ is a linear subspace of $E^\perp$
and $\Delta$ a discrete subgroup spanning a transversal to $V$ in
$E^\perp$.

Often one finds it convenient to consider tilings instead of
patterns.
The vertices of the canonical projection method tiling form the
projection pattern with canonical $K$. A clearer picture of the
construction of canonical projection method tilings arises from a
formulation based on dualization \cite{KramerSchlottmann89}. Here
one starts with an $N$-dimensional periodic polyhedral
complex 
and the tiles of the tiling consist of the projection of those
$d$-faces of the complex which satisfy an acceptance condition. To
formulate this condition one needs a dual complex. In particular,
there is a duality involved here, and if one interchanges the
complex with its dual one obtains the dual tiling.

For projection patterns we have a very explicit description of the
hull and of the dynamical system $(X,\Z^d)$ which we now describe.
Let $\Gamma= \pi^\perp(\Lambda)\cap V$. It acts on $V$ by
translation. Further let $S=\partial K+\Gamma\subset V$, the orbit
of the boundary $\partial K$ of $K$ under $\Gamma$. The rather
simple dynamical system $(V,\Gamma)$ extends to a dynamical system
$(\overline{V},\Gamma)$ which coincides with the old one on the
dense $G_\delta$-set $V\backslash S$. $\overline{V}$ is locally a
Cantor set and obtained from $V$ upon disconnecting it along the
points of $S$. In the most interesting cases, the set $S$ can be
described as follows: there is a finite set $\W$ of (affine)
hyperplanes of $V$ such that $S$ is the union of their translates
under the natural action of $\Gamma$. We call these planes and
their translates singular planes. In the canonical case, $\W$
consists simply of the spaces spanned by the boundary faces of the
acceptance domain $K$, i.e.\ $S=\bigcup_{w\in\W,x\in\Gamma}(W+x)$.
In general, $S$ may not be such a union but we have to restrict
our attention to that case. (In the formulation based on
dualization one can also identify acceptance domains, these are
the projections of the duals of the $d$-faces onto $E^\perp$.)
Finally we obtain $(X,\Z^d)$ from $(\overline{V},\Gamma)$ upon
splitting $\Gamma=\Z^\dpe\oplus \Z^d$ in such a way that $\Z^\dpe$
spans $V$ and setting $X=\overline{V}/\Z^\dpe$. Then $\Gamma$
induces an action of $\Gamma/\Z^\dpe\cong\Z^d$ on $X$. We
summarize this in the following set up.
\begin{setup}\label{set}
We consider data $(V,\Gamma,\W)$, a dense lattice $\Gamma$ of
finite rank in a euclidian space $V$ with a finite family
$\W=\{W_i\}_{i}$ of affine hyperplanes whose normals span $V$. We
make the additional assumption that these normals form an
indecomposable set in the sense that $\W$ cannot be written as a
union $\W=\W_1\cup \W_2$ such that the normals of $\W_i$ span
complimentary spaces $V_i$. We then define the dynamical system
$(X,\Z^d)$ from $(\overline{V},\Gamma)$ upon splitting
$\Gamma=\Z^\dpe\oplus \Z^d$ as above.
\end{setup}

\section{Cohomology groups for projection point patterns}

The cohomology groups $H^*(\Z^d,C(X,\Z))$ depend on the geometry
and combinatorics of the intersections of the singular planes,
i.e.\ of (affine) subspaces of the form $$ \bigcap_{(W,x)\in
A}(W+x)$$ where $A$ is some finite subset of $\W\times\Gamma$. We
call such a space a singular $l$-space if its dimension is $l$.
Let $\JJ_l$ be the set of singular $l$-spaces and denote the orbit
space under the action by translation $I_l:=\JJ_l/\Gamma$. The
stabilizer $\{x\in\Gamma|\hat\Theta+x=\hat\Theta\}$ of a singular
$l$-space $\hat\Theta$ depends only on the orbit class $\Theta\in
I_l$ of $\hat\Theta$ and we denote it $\Gamma^\Theta$. Fix
$\hat\Theta\in \JJ_{k}$, $l<k<\dpe$ and let
$\JJ_l^{\hat\Theta}:=\{\hat\Psi\in\JJ_l|{\hat\Psi}\subset
{\hat\Theta}\}$. Then $\Gamma^\Theta$ ($\Theta$ the orbit class of
$\hat\Theta$) acts on $\JJ_l^{\hat\Theta}$ and we let
$I_l^{\hat\Theta}=\JJ_l^{\hat\Theta}/\Gamma^\Theta$. We can
naturally identify $I_l^{\hat\Theta}$ with $I_l^{\hat\Theta'}$ if
$\hat\Theta$ and $\hat\Theta'$ belong to the same $\Gamma$-orbit
and so we define $I_l^{\Theta}$, for the class $\Theta\in I_{k}$.
$I_l^{\Theta}\subset I_l$ consists of those orbits of singular
$l$-spaces which have a representative that lies in a singular
space of class $\Theta$. Finally we use the notation $$ L_l =
|I_l|,\quad L^\Theta_l = |I^\Theta_l|.$$

\begin{theorem} Given data $(V,\Gamma,\W)$ as in \ref{set}.
If $L_0$ is finite $H^p(\Z^d,C(X,\Z))$ is a finitely generated
free abelian group, i.e.\ $H^{d-p}(\Z^d,C(X,\Z))\cong\Z^{D_p}$ for
finite $D_p$. If $L_0$ is infinite then $H^d(\Z^d,C(X,\Z))$ is
infinitely generated.
\end{theorem}
For better comparison with \cite{FHKmemoir} we wrote $$D_p= \rk
H^{d-p}(\Z^d,C(X,\Z)).$$ It is easily seen that $D_p= 0$ for $p<0$
or $p>d$.
\begin{theorem} Given data $(V,\Gamma,\W)$ as in \ref{set}.
If $L_0$ is finite the rank of the stabilizer $\Gamma^\Theta$
depends only on the dimension $\dim\Theta$ of the plane it
stabilizes, i.e.\ $$\rk\,\Gamma^\Theta=\nu\dim\Theta$$ where
$\nu=\frac{\mbox{\rm\footnotesize rank}\,\Gamma}{\dim V}$. In
particular, $\nu$ is a natural number.
\end{theorem}
The Euler characteristic is defined as $$e:=\sum_{p} (-1)^p
D_{p}.$$ We can determine it for arbitrary codimension. Define a
{\it singular sequence} to be a (finite) sequence $c = \Theta_1,
\Theta_2, ..., \Theta_k$ of $\Gamma$-orbits of singular spaces
strictly ascending in the sense that $\Theta_j \in
I^{\Theta_{j+1}}_{\dim \Theta_j}$, $\dim \Theta_j<\dim
\Theta_{j+1}$, and $\dim \Theta_1=0$. The length of the chain $c$
is $k$, written $|c| = k$.
\begin{theorem} Given data $(V,\Gamma,\W)$ as in \ref{set} with $L_0$
finite. Then the Euler characteristic is $$ e = \sum
(-1)^{|c|+\dim V}$$ where the sum is over all singular chains $c$.
\end{theorem}
Finally we present explicit formulae for the ranks $D_p$ in case
the codimension is smaller or equal to $3$. If $\{M_i\}_{i\in I}$
is a family of submodules of some bigger module we denote by
$\erz{M_i:i\in I}$ their span. For a finitely generated lattice
$G$ we let $\Lambda G$ be the exterior ring it generates.
\begin{theorem} Given data $(V,\Gamma,\W)$ as in \ref{set} with $L_0$
finite. We have \begin{itemize} \item[$\dim V=1$]
 $$ D_p = {\left(\nu \atop p+1
\right)},\quad p>0 $$ $$ D_0  = (\nu -1) + e, $$ \item[$\dim V=2$]
$$ D_p  = \left(2\nu \atop p+2 \right) + L_1\left(\nu \atop p+1
\right)-r_{p+1}-r_p,\quad p>0, $$ $$ D_0 = \left(2\nu \atop 2
\right)-2\nu+1 + L_1(\nu -1) + e -r_1 $$ where
$r_p=\rk\erz{\Lambda_{p+1}\Gamma^{\alpha}:\alpha\in I_1}$,
\item[$\dim V=3$] $$ D_p
 = \left( 3\nu\atop p+3\right)
+L_2 \left( 2\nu\atop p+2\right) +\tilde L_1\left( \nu\atop
p+1\right) - R_p - R_{p+1},\quad p>0  $$ $$ D_0
 = \sum_{j=0}^3 (-1)^j
\left( 3\nu\atop 3-j\right)+ L_2\sum_{j=0}^2 (-1)^j \left(
2\nu\atop 2-j\right)
+ \tilde L_1\sum_{j=0}^1 (-1)^j
\left(\nu\atop 1-j\right) + e  - R_1 $$ where $\tilde L_1 =
-L_1+\sum_{\alpha\in I_2}L_1^\alpha$ and
$$R_p=\rk\erz{\Lambda_{p+2}\Gamma^{\alpha}:\alpha\in I_2}
-\rk\erz{\Lambda_{p+1}\Gamma^{\Theta}:\Theta\in I_1}
+\sum_{\alpha\in I_2}
\rk\erz{\Lambda_{p+1}\Gamma^{\Theta}:\Theta\in I_1^\alpha} .$$
\end{itemize}
\end{theorem}

\section{Examples}

The above formulae for the ranks of the cohomology groups can be
evaluated with a computer using a derivative of a program to
compute Wyckoff positions of crystallographic space groups
\cite{EGN97}. It was already used to calculate the cohomology
groups of codimension~$2$ tilings \cite{GaKe99}. Franz G\"ahler
used it lately to calculate these groups for codimension~$3$
tilings, cf.\ Table~\ref{eqtable}. Apart from the first example,
the Ammann-Kramer tiling, these results have not been published
and the authors thank Franz G\"ahler for his permission to present
them here.

The tilings we look at here belong to a collection of icosahedral
tilings which is described in \cite{KramerPapadopolos94} and we
refer the reader for details and notation to that article. The
first three tilings of Table~\ref{eqtable} are obtained by the
variant of the cut and projection method which is based on
dualization. The fourth tiling, the Danzer tiling, has originally
been defined by a substitution \cite{Danzer89}. It is equivalent
to a tiling of Socolar and Steinhardt \cite{SoSt}. What is
important here is that the hull of Danzer's tiling can also be
described by data $(\Gamma,V,\W)$. The lattice $\Gamma$ is the
icosahedral projection
of $\Z^6$ in the first case and of the root lattice $D_6$ in the
three others. In particular, $V$ and the tiling are in all cases
both $3$-dimensional.

The Ammann-Kramer tiling, $\T^{(P)}$, is the canonical projection
method tiling obtained from the integer lattice $\Z^6$. Its
acceptance domain (the icosahedral projection of the unit cube) is
the triacontrahedron. We refer to \cite{FHKmemoir} for a
description of the singular spaces. This tiling is sometimes
referred to as $3$-dimensional Penrose tiling. Since the Voronoi
complex of $\Z^6$ and its dual are identical (up to a shift) there
is only one tiling of $P$-type.

The second tiling of Table~\ref{eqtable} and its dual (the third)
are derived from $D_6$. For the canonical $D_6$-tiling,
$\T^{(2F)}$, the singular planes are the lattice planes which are
perpendicular to the $3$-fold and $5$-fold axes of the lattice
whereas the singular planes for the dual canonical $D_6$-tiling,
$\T^{*(2F)}$, are those perpendicular to the $2$-fold axes.

The Danzer tiling can be locally derived from the canonical
$D_6$-tiling. Its relevant data differ from the latter tiling only
in the set of singular planes. These consist for the Danzer tiling
only of the lattice planes which are perpendicular to the $5$-fold
axes. Note that this tiling has surprisingly small cohomology
groups.

\begin{table}[ht]
\caption{Cohomology groups of various icosahedral tilings. Also
the quantities which enter into Theorem~5.4 are given.
}\label{eqtable}
\renewcommand\arraystretch{1.5}
\noindent
\begin{tabular}{|l|l|l|l|c|c|c|c|c|c|c|c|c|c|}
\hline Tiling &   $H^0$ & $H^1$ & $H^2$ & $H^3$ & $L_0$ & $e$&
$L_1$ & $\tilde L_1$ & $L_2$ & $R_1$ & $R_2$\\ \hline \hline
Ammann-Kramer & $\Z$ & $\Z^{12}$ & $\Z^{71}$  & $\Z^{180}$ & $32$
&$120$ & $46 $ & $74$ & $15 $ & $69$& $9$ \\ \hline  canonical
$D_6$& $\Z$ & $\Z^{13}$ & $\Z^{72}$ & $\Z^{205}$ & $56$ &
$145$&$45$ & $75$ & $16 $ & $73$& $9$ \\ \hline dual canon.\ $D_6$
& $\Z$ & $\Z^{12}$ & $\Z^{101}$ & $\Z^{330}$ & $64$ & $240$&$76 $
& $104$ & $15 $ & $69$& $9$
\\ \hline Danzer & $\Z$ & $\Z^{7}$ &
$\Z^{16}$ & $\Z^{20}$ & $1$&$10$ & $15$ & $15$ & $6 $ & $33$& $5$
\\
\hline
\end{tabular}
\end{table}

\providecommand{\bysame}{\leavevmode\hbox
to3em{\hrulefill}\thinspace}

\end{document}